\documentclass[aps,prb,preprint,draft,showpacs]{revtex4}

\def\figsize{13cm}

\usepackage{float}
\usepackage[final]{graphicx}
\usepackage{dcolumn}
\usepackage{bm}

\newcommand{\mm}[1]     {\ifmmode {#1} \else{}${#1}$\fi}
\newcommand{\mmm}[1]    {\ifmmode{}#1 \else{}${#1}$\fi}
\newcommand{\beq}[1]{\begin{equation}\label{#1}}
\newcommand{\eeq}{\end{equation}}
\def \LPCM{${\rm (La_{1-y}Pr_y)_{0.7}Ca_{0.3}MnO_3}$}
\def \lpcm{${\rm (La_{1-y}Pr_y)_{0.7}Ca_{0.3}MnO_3}$}
\def \pcso{${\rm Pr_{0.65}(Ca_{y}Sr_{1-y})_{0.35}MnO_3}$}
\def \lcm{${\rm La_{0.8}Ca_{0.2}MnO_3}$}
\def \lcm_ce{${\rm La_{0.5}Ca_{0.5}MnO_3}$}
\def \pcm{${\rm Pr_{0.7}Ca_{0.3}MnO_3}$}
\def \rsm{${\rm R_{0.5}Sr_{0.5}MnO_3}$}
\def \pcmo{${\rm Pr_{1-x}Ca_{x}MnO_3}$}
\def \lmo{${\rm LaMnO_3}$}
\def \rA{{\mm{\langle r_A\rangle}}}

\def\Os{${\rm ^{16}O}$}
\def\Oe{${\rm ^{18}O}$}
\newcommand{\ket}[1]{\left| #1\right\rangle}

\begin{document}

\title{\large Effect of oxygen isotope substitution and crystal
micro-structure on magnetic ordering and phase separation in {$\rm
(La_{1-y}Pr_{y})_{0.7}Ca_{0.3}MnO_3$}}


\author{V.~Yu.~Pomjakushin}
\affiliation{Laboratory for Neutron Scattering, ETHZ \& PSI, CH-5232
Villigen PSI, Switzerland}
\author{D.~V.~Sheptyakov}
\affiliation{Laboratory for Neutron Scattering, ETHZ \& PSI, CH-5232
Villigen PSI, Switzerland}
\author{K.~Conder}
\affiliation{Laboratory for Developments and Methods, PSI, CH-5232
Villigen PSI, Switzerland}
\author{E.~V.~Pomjakushina}
\affiliation{Laboratory for Developments and Methods, PSI, CH-5232
Villigen PSI, Switzerland}
\affiliation{Laboratory for Neutron Scattering, ETHZ \& PSI, CH-5232
Villigen PSI, Switzerland}
\author{ A.~M.~Balagurov}
\affiliation{Frank Laboratory of Neutron Physics, JINR, 141980, Dubna,
Russia}


\date{\today}

\begin{abstract}

The crystal and magnetic structures of the specified CMR manganite
system have been studied as a function of $y=(0.2-1)$ across the
metal-insulator (MI) transition, and of the oxygen mass ($^{16}$O,
$^{18}$O). We quantitatively show how the polaronic narrowing of the
carrier bandwidth and the crystal lattice micro-strains control the
volume fractions of the mesoscopic ferro- and antiferromagnetic
clusters. A well-defined dip in the transition temperatures and the
suppression of all the types of long range ordering seen near the
MI-transition at $y\simeq0.9$ indicate a key role of the quenched
disorder for the formation of the long-scale phase separated state.

\end{abstract}


\maketitle


\section{Introduction}
The presence of large \cite{zhao96} or giant oxygen isotope effects
\cite{zhao97,babushkina_nature} in the colossal magnetoresistance
(CMR) manganese oxides ${\rm A_{1-x}A'_xMnO_3}$ (A is rare earth
element, A$'$ is Ca, Sr, Ba) is now well established and it is
proven to be an intrinsic property of these systems.  Due to the
strong electron-phonon coupling mediated by the Jahn-Teller effect
the substitution of \Os\ by \Oe\ can significantly change the
effective exchange interaction between neighboring Mn ions
\cite{gorkov04}. In case of double exchange (DE) type of interaction
the increase in the oxygen mass strongly reduces the effective
charge carrier bandwidth, thus shifting the system towards
insulating state, while the standard superexchange with large
Hubbard energy gap is not supposed to be affected by the oxygen
mass. Thus the value of the isotope effect on the metallic and
insulating states of the above systems can reveal the underlying
interatomic and electron lattice interactions responsible for
particular magnetic, charge and orbital ordering, which are strongly
coupled together because of the 3d-electron anisotropy and the mixed
${\rm Mn^{3+}/Mn^{4+}}$ valence.  The low bandwidth manganite family
\lpcm\ (LPCM hereafter) has the fixed optimal hole doping $x=0.3$
and variable A-cation radius \rA\ that is linearly connected with
the Pr-concentration $y$. The principal effect of decreasing \rA\ is
a decrease in Mn-O-Mn bond angle, which leads to the decrease in the
electron transfer integral between the Mn-ions. The metal-insulator
(M-I) boundary lays at the Pr-concentration between $y=0.86$ and
$y=1.0$~\cite{hwang95,babushkina99}. The metallic state is
conditioned by the well known double exchange and it is
ferromagnetic (FMM), e.g. for $y=0$ \cite{wollan55,huang98}. The
magnetic structure of the insulating state can be both
antiferromagnetic (AFMI) with the pseudo-CE type magnetic structure
\cite{cox98,Bala01LPCM_magnetic} and ferromagnetic for the
Pr-concentrations close to $y=1$.  The AFMI state is a Mott
insulator state with the charge ordering of the ${\rm
Mn^{3+}/Mn^{4+}}$ ions similar to the case of \lcm_ce, whereas the
origin of the genuine ferromagnetic insulating phase (FMI) is not
completely understood and it is characterized by a non-ordinary kind
of orbital/charge ordering \cite{fisher03, papavassiliou03,
kajimoto04,pissas05}.  The LPCM system can be converted from the
metallic to insulating state by the isotope substitution
\cite{babushkina_nature,bala99:isotope}.  Similar giant isotope
effect was observed in ${\rm
(La_{0.5}Nd_{0.5})_{0.67}Ca_{0.33}MnO_3}$ \cite{zhao97}. The
complete suppression of the ferromagnetic transition temperature
cannot be attributed to a solely polaronic narrowing of the
bandwidth within the DE model \cite{zhao96,gorkov04}, but requires
an inhomogeneous phase separated state, which was indeed directly
observed by different experimental techniques in \lpcm\
\cite{Bala01LPCM_magnetic,tokunaga04} and in the similar series $\rm
La_{5/8-y}Pr_yCa_{3/8}MnO_3$ \cite{uehara99,littlewood99}. The
neutron diffraction study \cite{bala99:isotope} has shown that the
increase in the oxygen mass completely converts the dominating FMM
metallic phase into the insulating charge ordered AFMI one in LPCM
with $y=0.75$. The presence of the long scale percolate phase
separation in the manganites itself has been attracting special
experimental and theoretical attention. There are two main concepts
of the mesoscopically inhomogeneous state in the manganites. One
standpoint is that the intrinsic quenched disorder enhances the
fluctuations of the competing orders near the original bi-critical
point \cite{burgy01,burgy04,alonso02,blake02,akahoshi03,deteresa05}.
In another approach, the lattice distortions and the long-range
strain \cite{littlewood99, ann04, podzorov01, sharma05} are the
dominant factors controlling the phase separation, similar to the
one observed at the martensite type structural transition. Motivated
by the spectacular giant isotope effect and the problem of phase
separation we have undertaken a systematic neutron and synchrotron
x-ray diffraction study of a series of the samples \lpcm\ for
$y=0.2-1.0$ with \Os\ and \Oe.  In the paper we present the detailed
data on the structure and magnetic properties of the FMM, FMI and
the charge ordered AFMI phases and discuss the key parameters
conditioning the giant isotope effect and the phase separation in
the LPCM-system. Earlier we have performed a diffraction study with
another series of samples of the same system prepared by a different
chemical route~\cite{Bala01LPCM_magnetic}, in which the effect of
the oxygen isotope substitution has been studied only for one
concentration $y=0.75$. Moreover the different synthesis route used
in the present study gives different microstructure parameters,
which as we show below, have influence on the phase separation and
hence on the oxygen substitution effect.

The paper is organized as follows. In Sec. \ref{exp} the details of
the sample synthesis, oxygen treatment and the experimental
measurements are given. In Sec. \ref{magn_sus} the temperature
dependences of the magnetic susceptibility $\chi(T)$ are analyzed.
The ferromagnetic transition temperatures $T_C$ determined from
$\chi(T)$ are used as independent estimations of $T_C$ complementary
to ones determined from the neutron diffraction measurements. In
Sec. \ref{crys_str} the room temperature crystal structure and
microstructure parameters as a function of $y$ are compared for \Os\
and \Oe-samples. In Sec. \ref{OO} the structure parameters are
analyzed as a function of temperature between 15~K and 1200~K across
the orbital order-disorder transition at $\simeq700$~K. In
Sec.~\ref{CO} we show the evolution of the orbital and charge order
as a function of doping $y$. In Sec. \ref{mag_state} the data on the
temperature dependence of the magnetic state and the spatial phase
separation are presented. The magnetic transition temperatures and
the magnetic ground state (the phase fractions, ordered magnetic
moments) are obtained as a function of doping $y$ and the oxygen
mass. Sec.~\ref{u_str} explains the connection between the phase
separation and the lattice micro-strains (static variance of metric
parameters). In Sec.~\ref{qcp} the suppression of all types of
ordering near critical concentration $y_c=0.9$ is discussed. This
effect can be well interpreted in the theoretical model
\cite{burgy01,burgy04}, which assumes the presence of compering
states and quenched disorder near $y_c$.

\section{Samples. Experimental}
\label{exp}

The crystal structure parameters of a LPCM sample for given $y$ depend
not only on $y$ but also on the sample preparation procedure. For
example, the spontaneous orthorhombic strain, thermal displacement
parameters (reflecting also the local static disorder), the lattice
micro-strains decrease as a function of annealing time and of the
cooling time through the high temperature pseudocubic-orthorhombic
structural transition. Thus, a precise study of a real LPCM system as
a function of $y$ and the oxygen mass can be done only within a series
of samples prepared and thermally treated at the very identical
conditions.  The samples, which are denoted as ``O-series'', were
calcined at temperatures 1270-1570~K for 100h from ${\rm La_2O_3}$,
${\rm Pr_6O_{11}}$, ${\rm MnO_2}$ oxides and ${\rm CaCO_3}$ with 3
intermediate grindings. These samples we call ``as-prepared''. The
final \Oe- and \Os-samples were obtained via respective oxygen isotope
exchange in closed quartz tubes in parallel under the controlled gas
pressure slightly above 1 bar at 1270~K during 40h and then cooled
down to 290~K with the rate 30 K/h. The \Oe\ -samples had 80\% of
\Oe-isotope, measured by the weight gain after the oxygen
exchange. The control weighing of the \Os-sample gave the same mass
within accuracy $0.03$\%. The mass of each sample was about 2~g. The
oxygen content in all the samples was determined by the
thermogravimetric hydrogen reduction \cite{conder02} and amounted to
3.003(5). The samples with $y=0.85$ and 0.95 were prepared only with
\Os. One of the samples which has been studied earlier
\cite{bala99:isotope,Bala01LPCM_magnetic,Bala01LPCM_crystal}, was also
oxygen treated in the same way as all the samples from the
O-series. This sample is from a series prepared by a synthesis from
the aqueous solutions of the respective metals nitrates (denoted as
``N-series'').

The $ac$ magnetic susceptibility $\chi(T)=\chi'(T)+i\chi''(T)$ was
measured in zero external field with amplitude of the $ac$ field
10~Oe and frequency 1~kHz using Quantum Design PPMS station. Neutron
powder diffraction experiments were carried out at the SINQ
spallation source of Paul Scherrer Institute (Switzerland) using the
high-resolution diffractometer for thermal neutrons HRPT
\cite{Fischer00HRPT} ($\lambda=1.866, 1.494$~\AA, high intensity
mode $\Delta d/d\geq1.8\cdot10^{-3}$), and the DMC diffractometer
\cite{dmc} situated at a supermirror coated guide for cold neutrons
at SINQ ($\lambda=2.56$~\AA). All the temperature scans were carried
out on heating. X-ray synchrotron diffraction at room temperature
was done at the Material Sciences beam line (MS, SLS/PSI). The
refinements of the crystal and magnetic structure parameters were
carried out with {\tt FULLPROF}~\cite{Fullprof} program, with the
use of its internal tables for scattering lengths and magnetic form
factors.

\section{Results and discussion}

\subsection{Magnetic susceptibility}
\label{magn_sus}

\begin{figure}[p]
\includegraphics[width=\figsize]{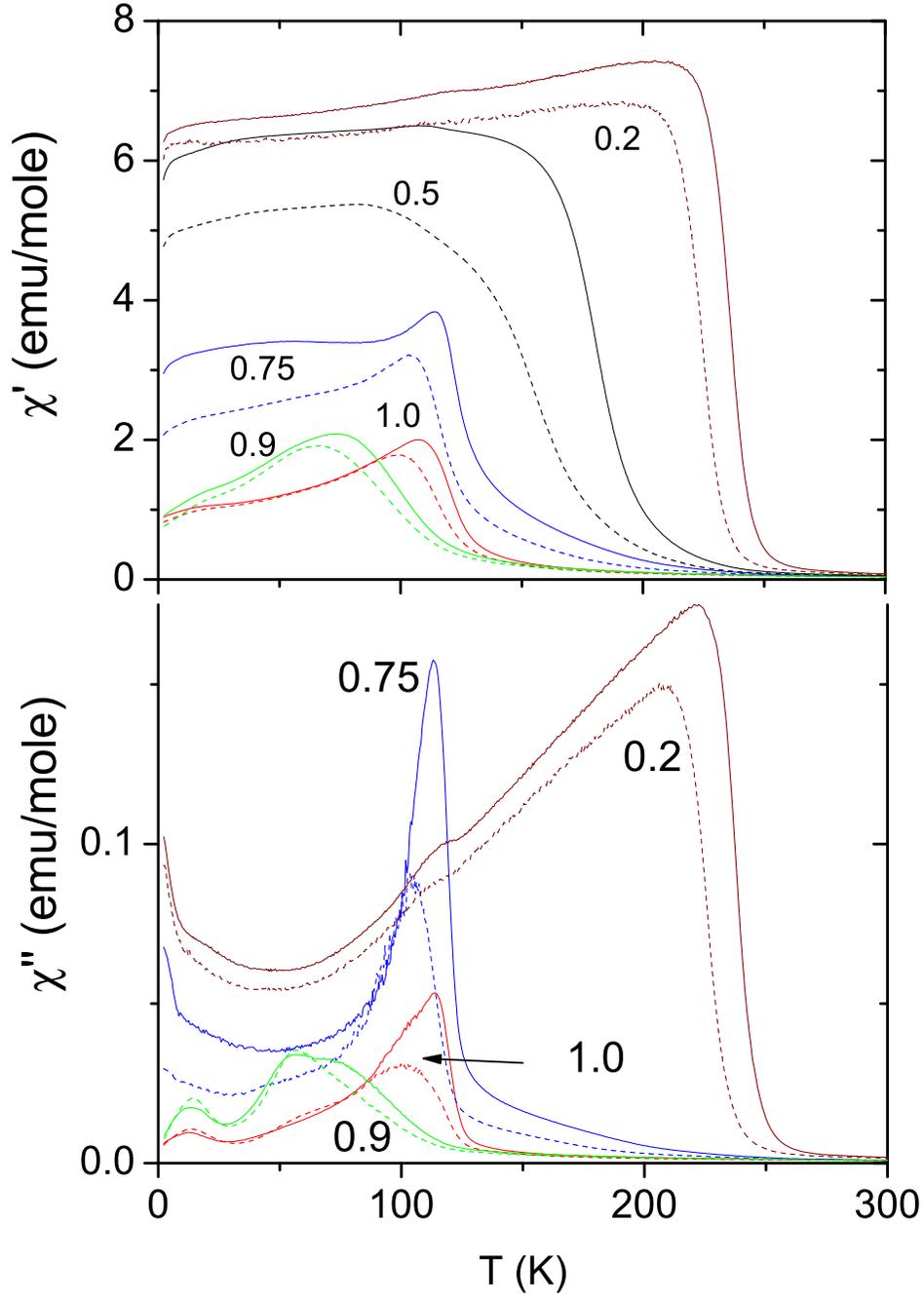}
\caption{(color online) Real $\chi'$ and imaginary $\chi''$ parts of
the $ac$ magnetic susceptibility are shown as a function of
temperature for several samples with the Pr-content $y$ indicated in
the plots. \Os\ and \Oe-samples are shown by solid and dashed lines
respectively. $\chi''$ for $y=0.5$ (which is similar to one for
$y=0.2$) is not shown to avoid plot overload.} \label{chi(T)}
\end{figure}

\begin{figure}[p]
\includegraphics[width=\figsize]{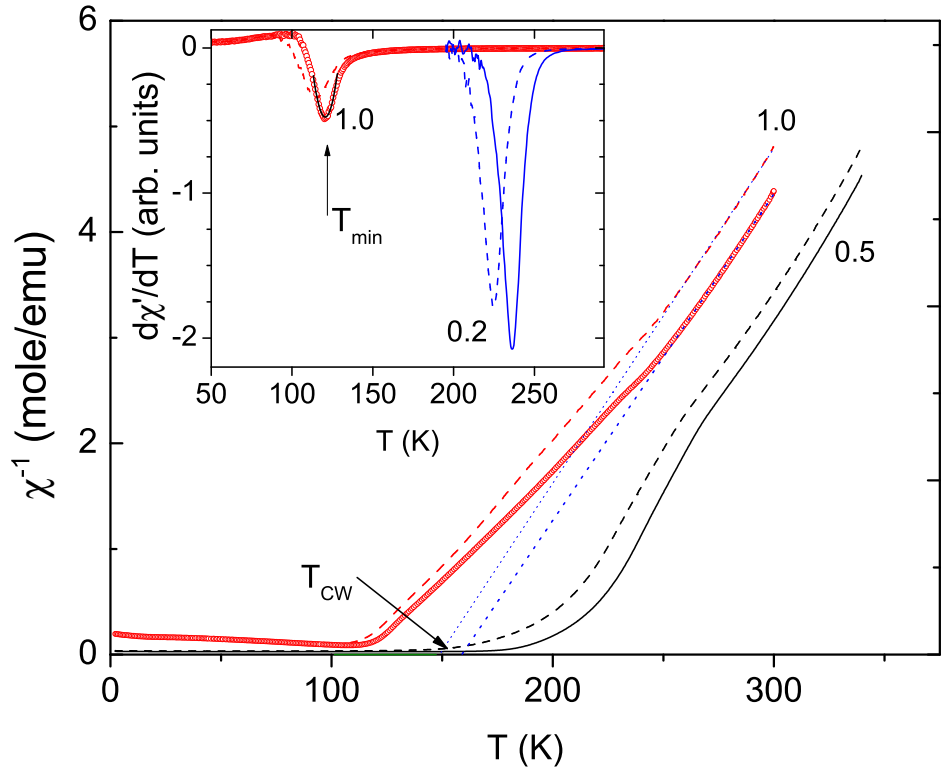}
\caption{(color online) $\chi'^{-1}$ as a function of temperature
for the Pr-content $y$ indicated in the plots. The dotted lines show
the fits of high temperature part of $\chi'$ to the Curie-Weiss low.
The refined Curie temperatures $T_{CW}$ are indicated. The insert
shows two examples of the first derivative $d\chi'/dT$ as a function
of temperature.  \Os\ and \Oe-samples are shown by solid and dashed
lines, respectively except for the y=1.0 (\Os) sample, which is
shown by circles. The solid line in the insert for this sample shows
example of the parabolic fit. The refined temperature of the minimum
is indicated in the insert.}

\label{chi-1(T)}
\end{figure}

The main use of the magnetic susceptibility data $\chi(T)$ for the
purposes of the present work is the determination of the ferromagnetic
transition temperatures, complementary to the NPD data.  The $\chi(T)$
as a function of temperature is shown in Fig.~\ref{chi(T)} for several
selected samples [since the $\chi(T)$ are quite similar for the
samples with close values of Pr-concentrations, not all the
$\chi(T)$-curves are shown to avoid the plot overload].  This plot
nicely illustrates the systematic decrease in the magnetic transition
temperatures as a function of the oxygen mass. All the samples show
(Fig.~\ref{chi-1(T)}) the ferromagnetic type of the Curie-Weiss
susceptibility $\chi'(T)$ at high temperatures. The slope of the
$\chi'^{-1}(T)$ becomes non-linear below $~250$~K due to the presence
of the antiferromagnetic correlations which are developed as the
temperature decreases down to the N\'{e}el temperature
$T_N\simeq150$~K.  The Curie-Weiss temperatures $T_{CW}$ were
determined from the fit of the high temperature susceptibility to
$\chi'(T)=C/(T-T_{CW})$ in the temperature range ($270-335$~K)
depending on $y$ content.  The Curie-Weiss temperatures amounted to
$T_{CW}=165-240$~K for the \Os-samples for $y=1-0.2$.  They
are in agreement with  the ferromagnetic transition temperatures $T_C$
determined from NPD only for $y=0.2$ and $0.5$, while for the higher
Pr-content $T_{CW}$ is significantly higher than $T_C$. In the
\Oe-substituted samples the $T_{CW}$ is decreased similar to the
change in $T_C$, implying smaller effective electron transfer
integral which mediates the ferromagnetic coupling for the heavier
oxygen ion. The Curie-Weiss constant $C=N_A 2S(S+1)\mu_B/3k$ allows
estimating the paramagnetic Mn-spin value $S$.  The experimental
values of $C$ amounted to $4.6-6$~emu/mol depending on the
$y$-content that correspond to the effective spin $2S\simeq5.2-6$, instead
of the expected value of 3.7 for the given $\rm Mn^{3+}/Mn^{4+}$
ratio. The enhanced value of the Curie-Weiss constant is a typical
feature of the ferromagnetic manganite system and is attributed to the
formation of the ferromagnetic clusters well above $T_C$
\cite{franck02,deteresa97}. The quantitative analysis of $C$ as a
function of $y$ is difficult because the Curie-Weiss constant value
depends on the temperature interval used for the fit and the large
variation of the $T_C$.  The imaginary part of the susceptibility
shows a maximum at a temperature very close to $T_C$ for all
$y$ (Fig.~\ref{chi(T)}). The maximum is caused by the absorption due to
the ferromagnetic domains formation with sufficiently short relaxation
times $\tau\sim1/(2\pi f)$ close to $T_C$, where $f$ is the external
field frequency.  This maximum is asymmetric and can be very broad, so
the temperature of the maximum cannot be used for quantitative estimation
of the $T_C$. A good way of the evaluation of the ferromagnetic
transition temperature is the analysis of the derivative of the
magnetic susceptibility $d\chi\prime/dT(T)$ which has a well
defined minimum as shown in the insert of Fig. \ref{chi-1(T)}.  The
temperature of the minimum $T_{min}$ will be taken as an independent,
complementary to the NPD data, estimation of $T_C$.

\subsection{Crystal structure and microstructure}
\label{crys_str}

\begin{figure}[p]
\begin{center}
\leavevmode
\includegraphics[width=\figsize]{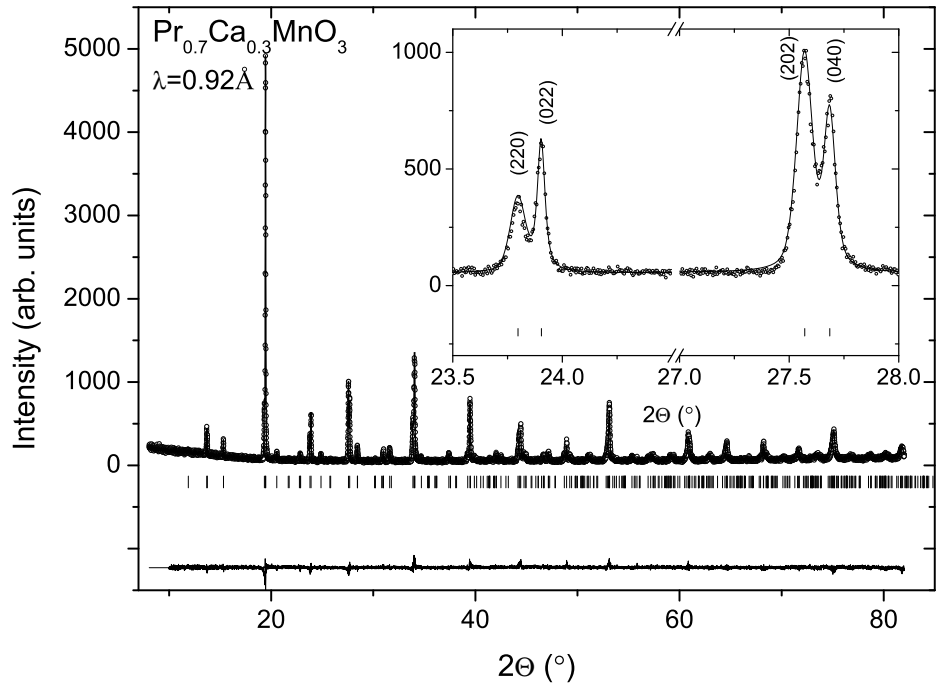}
\end{center}
\caption{An example of the Rietveld refinement pattern and difference
plot of x-ray synchrotron diffraction data (MSP/SLS, PSI, $\lambda=0.92$~\AA) for the
sample with y=1.0 at room temperature. The insert illustrates
anisotropic line broadening along $a$-axis.}

\label{SLSpatterns}
\end{figure}

\begin{figure}[p]
\includegraphics[width=\figsize]{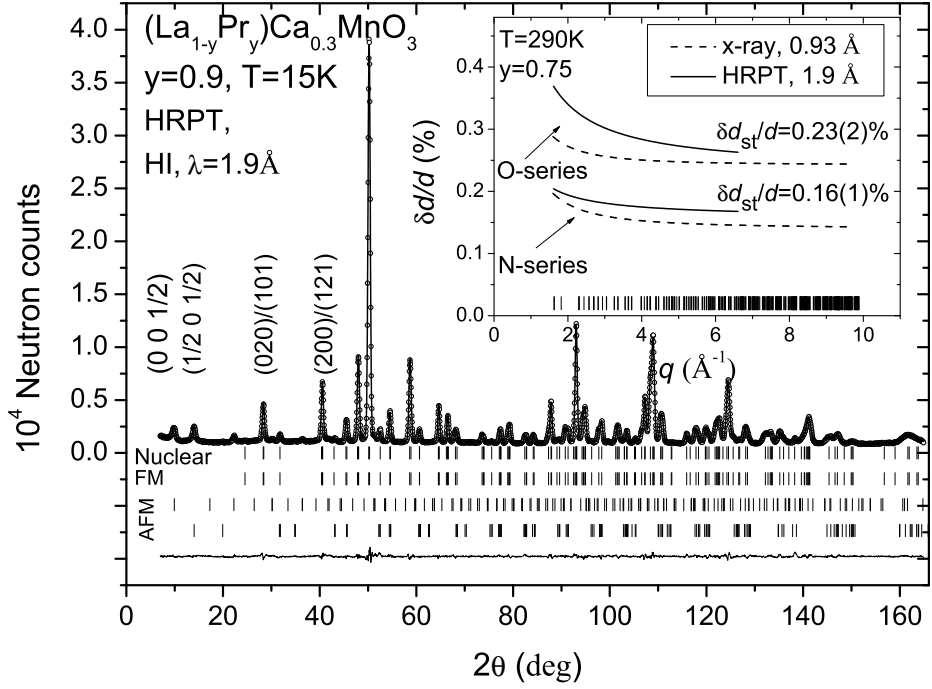}
\caption{An example of the Rietveld refinement pattern and
difference plot of the neutron diffraction data for the sample with
$y=0.9$ at T=15K. The most intensive magnetic peaks are indicated by
Miller indices.  The insert illustrates the reduced FWHM of the
Bragg peak broadening ($\delta d/d$) as a function of $q$ for two
samples with $y=0.75$ from O- and N-series at room temperature
measured by neutron and x-ray synchrotron diffraction. ($\delta
d/d$) was calculated from the refined FWHM-width of the Bragg peaks
after deconvoluted with the instrument resolution function. $\delta
d_{st}/d$ are the refined isotropic micro-strain value.  See the
text for details.}

\label{HRPTpatterns}
\end{figure}

\begin{figure}[p]
\includegraphics[width=\figsize]{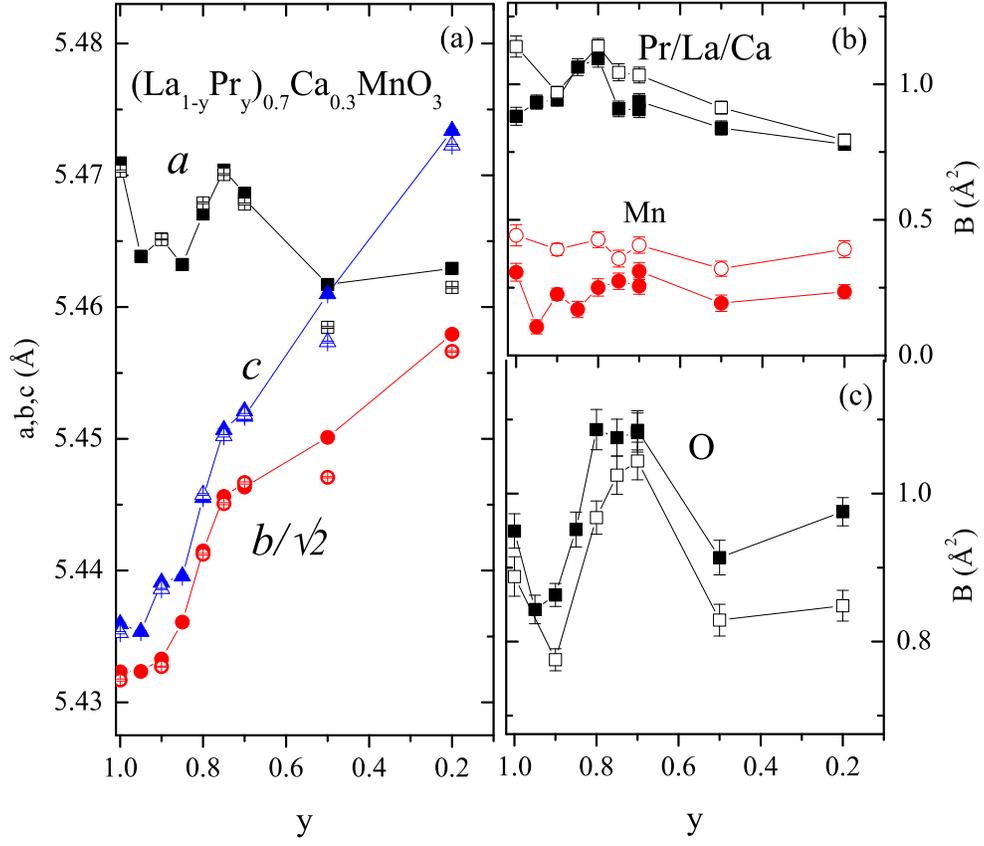}
\caption{(color online) (a) The lattice constants (sp. gr. $Pnma$) and
(b,c) isotropic thermal parameters at room temperature $T=290$~K.
\Os\ and \Oe-samples are shown by closed and open symbols
respectively.}
\label{Babc}
\end{figure}

\begin{figure}[p]
\includegraphics[width=\figsize]{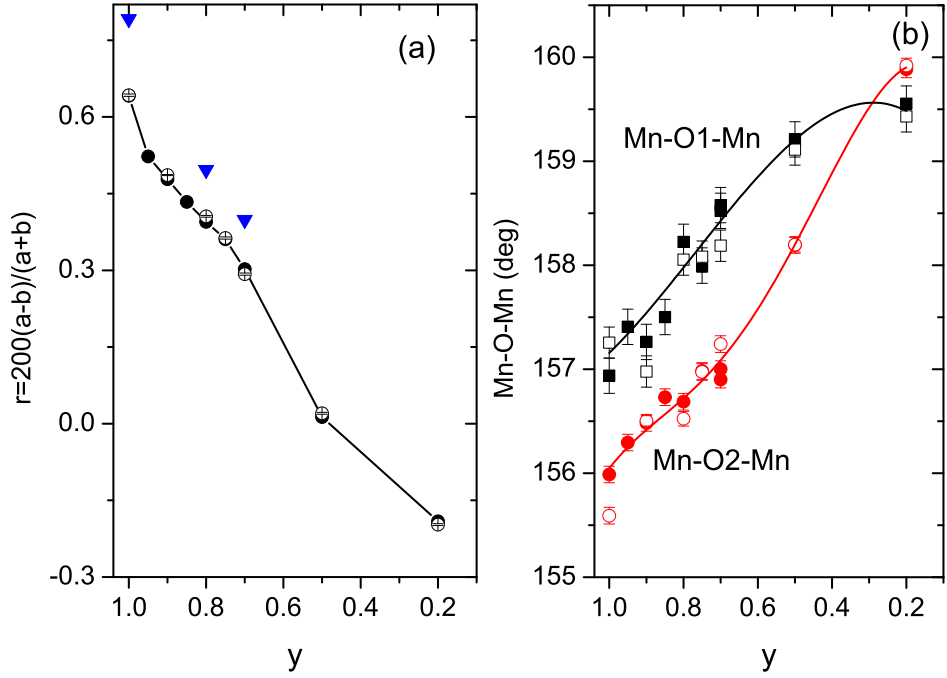}
\caption{(color online) (a) The orthorhombic strain $r$ and (b)
Mn-O-Mn bond angles as a function of Pr concentration (sp. gr.
$Pnma$) at room temperature $T=290$~K. \Os\ and \Oe-samples are
shown by closed and open symbols respectively. The triangles in (a)
show the orthorhombic strain in the ``as prepared'' samples for
comparison. The lines are guides for the eyes.} \label{r+angle}
\end{figure}

The crystal structure for all the compositions $y=0.2-1.0$ at all
temperatures is well refined in single phase in the orthorhombic
space group $Pnma$ with the standard for these compounds structure
model (see e.g. \cite{Bala01LPCM_crystal}). In general, the low
temperature state is spatially mesoscopically phase separated, but
the lattice constants of the phases are too close to each other and
the two phase refinement is not reliable. However, there are cases
where the structure separation is explicitly visible, e.g. in the
system \rsm\ ($\rm R=Sm, Td, Tb$) the metrics of the phases are
sufficiently different and both phases can be identified in the
neutron diffraction experiment\cite{bala05}.  The examples of the
x-ray synchrotron diffraction pattern and the neutron diffraction
pattern and their refinements are shown in Fig.~\ref{SLSpatterns}
and Fig.~\ref{HRPTpatterns}, respectively.  All the room temperature
($T=290$~K) crystal structure parameters except for the thermal
displacement parameter do not depend on the oxygen mass within the
accuracy 0.2\%.  The lattice constants, which are determined with
the best accuracy (Fig. \ref{Babc}(a)) are slightly systematically
smaller for the \Oe-samples: for $y\leq0.5$ the unit cell volume $V$
is 0.2\% smaller, for the $y>0.5$ the volumes $V$ match within
0.02\%.  The only structure parameter which is noticeably and
systematically changed by the oxygen isotope substitution is the
thermal displacement parameter $B$. The oxygen thermal parameter is
slightly decreased, while the Mn and A-cation ones are pronouncedly
increased as a function of oxygen mass (Fig. \ref{Babc}(b,c)).
These changes agree with the expected behavior of the $B\sim (k_B T)
\overline{\omega^{-2}}/M$, where $M$ is the atom mass, the
$\omega^{-2}$ is averaged over the phonon density of states
$Z(\omega)$~\cite{LoveseyI:W}. The parameter
$\overline{\omega^{-2}}$ should increase as a function of an average
atom mass provided that the elastic force constants are not changed.
Hence the thermal parameters of Mn and A-cation should increase,
while the oxygen $B$-value behavior cannot be simply predicted
because both $\overline{\omega^{-2}}$ and $M$ are changing.  Figure
\ref{r+angle} illustrates systematic distortion of the structure as
a function of the Pr-concentration $y$. The Mn-O-Mn bond angles are
systematically decreased conditioning the change from the metallic
to insulating state at the concentration above $\sim0.85$. At the
same time the orthorhombic strain $r$ is substantially increased
with the increase in $y$.  The values of strain $r$ for the
``as-prepared'' samples (also shown in the Fig.~\ref{r+angle}a) are
significantly larger stressing the point that the systematic study
of the present system as a function of $y$ and the oxygen mass is
only possible within a series of samples prepared and treated in
identical way.

The microstructure of the samples was obtained from the analysis of
the Bragg peak broadening (anisotropic in general) as a function of
scattering angle ($2\theta$) by the following procedure. The Bragg
peaks were described by the pseudo-Voight distribution function with
the Gaussian ($\Gamma_G$) and Lorentzian ($\Gamma_L$) components.
The total peak width $\Gamma$ was calculated from the individual
components using a 5th order polynomial approximation as described
in Ref.~\cite{thompson87} and implemented in the {\tt FULLPROF} code
\cite{Fullprof}. The angular dependence of the peak width was
approximated by the Cagliotti-type function
$\Gamma_G(2\theta)=((U\tan^2\theta+\sigma_G^2) + V\tan\theta + W
+\sigma_{1G}/\cos^2\theta)^{(1/2)}$ and
$\Gamma_L(2\theta)=(X\tan\theta+\sigma_L) +
(Y+\sigma_{1L})/\cos\theta$ for the Gaussian and Lorentzian
components respectively. All the widths correspond to the full width
at half-maximum (FWHM) and the angles are measured in radians. $U$,
$W$, $U$ and $Y$ were fixed to the values given by the instrumental
resolution function. $\sigma_G(2\theta)$ and $\sigma_L(2\theta)$ are
the Rietveld refined Gaussian and Lorentzian components related to
the sample micro-strains. $\sigma_{1G}$ and $\sigma_{1L}$ are the
refined components related to the finite size broadening. The total
widths both for the micro-strain ($\sigma(2\theta)$) and the finite
size ($\sigma_1$) broadening were calculated from the individual
Gaussian and Lorentzian refined components by means the above
mentioned 5th-order polynomial approximation. The apparent sizes
($L$) are connected with the broadening $\sigma_1$ by well-known
Debay formula $\sigma_1=\lambda/L$. The micro-strains or the static
fluctuations of crystal lattice constants are determined from the
variance of the metric parameters ($A$,$B$,$C$) of the reciprocal
space. The square of the reciprocal distance $1/ d$ for a Bragg
reflection with Miller indices ($hkl$) for the orthorhombic singony
reads $1/d^2=M_{hkl}=A h^2+B k^2 + C l^2$. The variance of $M_{hkl}$
is related to the widths of the diffraction peaks as
$\sigma(2\theta)=\sigma(M_{hkl})/M_{hkl}\tan\theta$. General
description of the anisotropic peak broadening is given in Ref.
\cite{stephens99}. In our case we consider the isotropic
micro-strains $(\delta d_{st}/d)$, where $d$ stands for  $a$, $b$ or
$c$ lattice constants, and an additional micro-strain along $a$-axis
$(\delta d_{st}/d)_{[100]}$ that have the contributions to the
reciprocal distance variance $\sigma(M_{hkl})=2 (\delta d_{st}/d)
M_{hkl}$ and $2(\delta d_{st}/d)_{[100]} A h^2$, respectively. The
presence of the anisotropic broadening is nicely illustrated by the
two doublets shown in the insert of Fig.~\ref{SLSpatterns}.

The sample related peak broadening is comparable with the resolution
of neutron diffraction instrument HRPT, therefore we have performed
a comparative ultra-high resolution synchrotron x-ray diffraction
study of several samples at room temperature. The examples of the
reduced sample related peak widths $\delta d/d=\Gamma_{\rm
sample}/\tan(\theta)$ as a function of scattering vector $q=2\pi/d$
are shown in the insert of Fig. \ref{HRPTpatterns}. The width
$\Gamma_{\rm sample}$ includes the isotropic micro-strains and the
finite size broadening components. One can see that the experimental
broadening is dominated by the micro-strain effect, which gives a
constant term to $\delta d/d(q)$, whereas the finite size effect
gives a $\propto 1/q$ term. The apparent sizes $L$ are larger than
$2\cdot10^3$~\AA\ giving a relatively small contribution to the
broadening (assuming $1/L=0$ increases the refined micro-strain
values by less than 10\%). The micro-strains refined from the
neutron and synchrotron data agree within the statistical error
bars. This implies that in spite of lower resolution of the neutron
diffraction data ($\delta d/d\simeq2\cdot10^{-4}$ and
$\geq1.8\cdot10^{-3}$ for the synchrotron x-ray and neutrons
respectively) we can reliably extract the microstructure parameters
from the NPD data. The lattice micro-strains allow us to
characterize the amount of quenched disorder and the presence of the
charge ordering as discussed in the following sections.

There are two transitions in \lpcm\ associated with the changes in the
crystal structure. The first transition at high temperature is a fist
order isostructural phase transition from the pseudo-cubic to the
orthorhombic orbital ordered structure similar to the one observed in
the stoichiometric $\rm LaMnO_3$ \cite{carvajal98}. The second
transition at lower temperature is related to the charge ordering of
the Mn-ions and is accompanied by the lowering symmetry to the
monoclinic one and the unit cell doubling along the orthorhombic
$a$-axis. We have studied the crystal structure of several
compositions of LPCM up to the high temperatures across the transition
to the pseudocubic phase.

\subsection{Orbital order-disorder transition}
\label{OO}

\begin{figure}[p]
\includegraphics[width=\figsize]{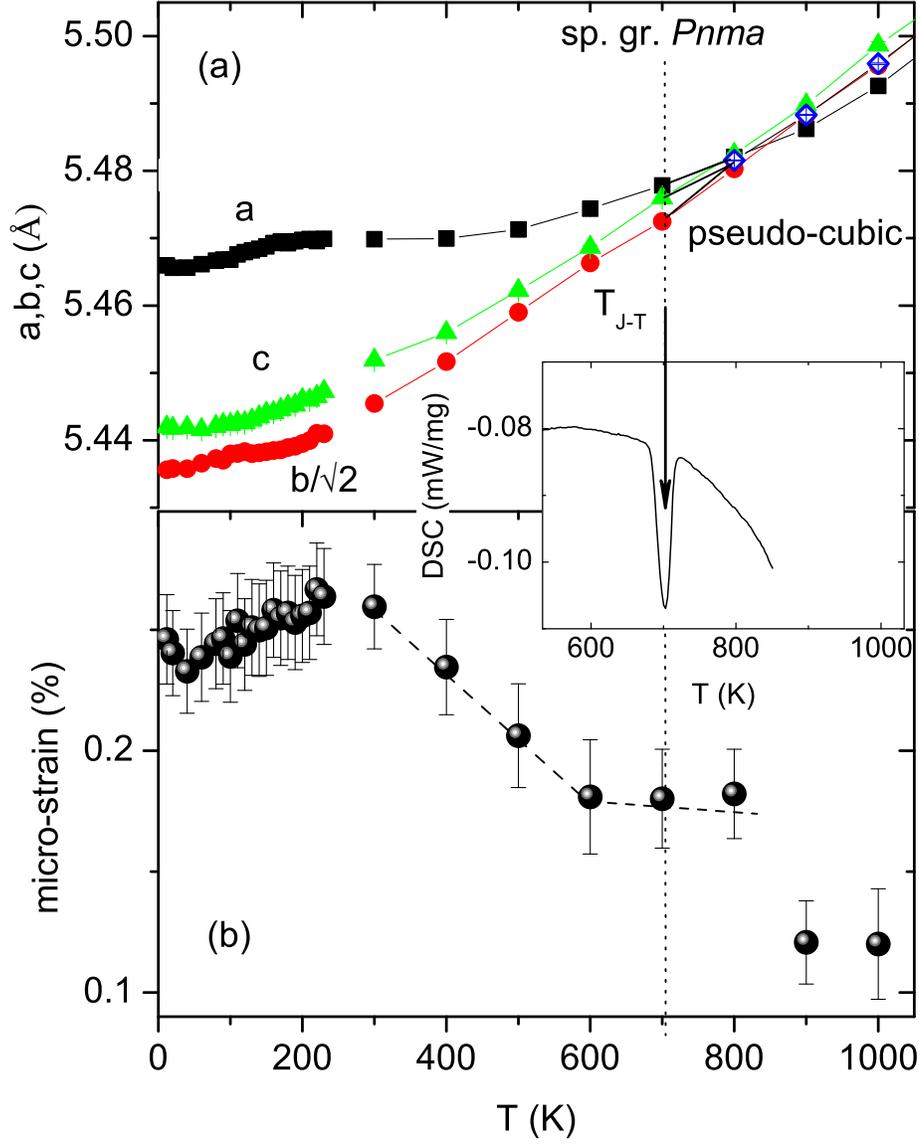}
\caption{(color online) (a) Lattice constants and (b) isotropic
micro-strains as a function of temperature in the \lpcm\ (y=0.75).
The insert shows the differential scanning calorimetry signal with
the same x-axis. The minimum indicate the presence of the latent
heat of the first order phase transition at the temperature
$T_{J-T}$.} \label{abchilow75}
\end{figure}

The study the crystal structure of LPCM up to the transition to the
pseudocubic phase allowed us to get a wide range of the changes in
the crystal structure, which helped to understand the temperature
behavior of the crystal structure at low temperature. Figure
\ref{abchilow75} shows the crystal lattice constants and the
isotropic micro-strain parameter $(\delta d_{st}/d)$ in the LPCM
sample with y=0.75. The latent heat at the transition temperature
$T_{J-T}$ (insert in Fig.~\ref{abchilow75}) points to the first
order phase transition. Above the transition the crystal structure
is metrically cubic and the refinement in the profile matching mode
in the cubic metrics gives the same quality of the fit ($\chi^2$)
with the lattice constant shown by rhombs in Fig.~\ref{abchilow75}a.
In addition, the $\rm MnO_6$ octahedron gets regular above the
transition.  The structure is thus becoming pseudo-cubic. This has
inspired us to search for a possible alternative cubic crystal
structure description, similar to the attempt undertaken in
\cite{carvajal98}. The diffraction pattern at 800 K is indexed in
the cubic cell with $a=7.76$~\AA. The list of possible space groups
with their numbers in brackets, which does not contradict the set of
observed diffraction peaks contains seven possibilities: $P2_13$
(198), $P4_23m$ (208), $P23$ (195), $Pm\bar{3}$(200), $P432$ (207),
$P\bar{4}32$ (215) and $Pm\bar{3}m$ (221). Our trial structure
solutions (carried out with the program FOX \cite{fox}) did however
fail to find a suitable candidate structure model. As it was
actually pointed out in \cite{carvajal98}, the failure could have
been predicted by the even earlier studies by Glazer
\cite{glazer75}, who did not deduce any primitive cubic space group
for the structure with lattice constant $2a_p$ for the distorted
perovskite which could have originated from the coherent octahedra
tilts.

The micro-strain value $(\delta d_{st}/d)$ is markedly increased
below the transition temperature (Fig. \ref{abchilow75}b).  The
lattice constant mismatch in the adjacent crystal domains and/or
twins naturally creates the internal micro-strains below the
transition $T_{J-T}$. This is important to note because the phase
separation at low temperature depends on the value of the
micro-strains, which will be inevitably present in the LPCM system
due to this structure transition.

\begin{figure}[p]
\includegraphics[width=\figsize]{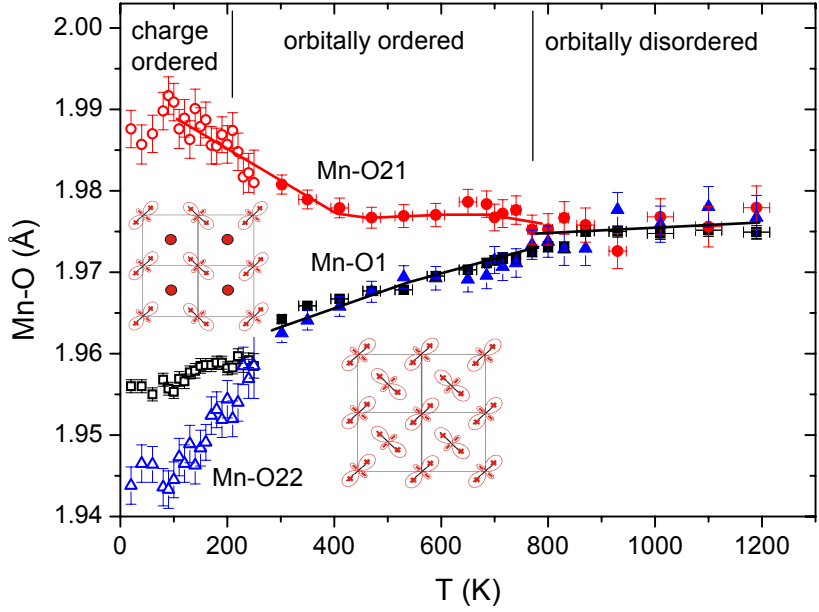}
\caption{(color online) Bond length Mn-O1 and Mn-O2 being approximately along and
perpendicular $b$-axis, respectively as a function of temperature in
\lpcm\ for y=0.7. The structure undergoes a change from the orbitally
disordered pseudocubic phase to the antiferrodistorsive type of
orbital ordering (OO) below 800K and to the charge ordered (CO) state
at lower temperatures. The sketches of the $z^2$-type Mn-orbitals in the
$(ac)$ plane are shown in the inserts.  In the OO phase each Mn-site
is statistically occupied by $\rm Mn^{3+}$ and $\rm Mn^{4+}$. In the
CO-phase the fully occupied $\rm Mn^{3+}$-sites are shown with
$z^2$-type orbitals, while the sites occupied by both $\rm Mn^{4+}$
and $\rm Mn^{3+}$ are indicated by the circles.  See the text for the
details. }
\label{70allT}
\end{figure}

Figure  ~\ref{70allT} shows the bond lengths between manganese and
oxygen atoms in the $\rm MnO_6$-octahedra as a function of
temperature. Mn-O1 is the bond directed to the oxygen in the
position shifted along $y$-direction with respect to the Mn-atom.
Mn-O21 and Mn-O22 are the bonds to the oxygen atoms located roughly
in the $(ac)$ plane. The transition at $T_{J-T}$ is a Jahn-Teller
type of transition connected with the lifting up of the $e_g$
orbital degeneracy. The filling of the $e_g$ orbitals
$\ket{e_g}=\cos(\theta/2)\ket{z^2}+\sin(\theta/2)\ket{x^2-y^2}$ can
be determined from the value of an angle $\theta$, which is the
polar angle in the plane of normal coordinates $(Q_2,Q_3)$
\cite{kanamori60}. The coordinates are given by the distortion of
the $\rm MnO_6$ octahedron $Q3=(1/\sqrt{2})(l-s)$,
$Q2=(2/\sqrt{6})(2m-l-s)$, where $s$, $m$, $l$ are the short, medium
and long Mn-O bond lengths, respectively. One can see from the
Fig.~\ref{70allT} that above $T_{J-T}$ all three bonds are equal,
implying that both $e_g$-orbitals are equally populated. Below the
transition two of them remain equal ($m=s$), that gives the angle
$\theta=2\pi/3$. This angle corresponds to the complete filling of
the $y^2$-orbital or for another choice of axes to equivalent $z^2$-
or $x^2$- orbitals [they are related to each other by $2\pi/3$
rotation in the $(Q_3,Q_2)$-plane]. In the $Pnma$ structure it
corresponds to the antiferrodistorsive type of orbital ordering (OO)
schematically shown in the Fig.~\ref{70allT}. It is interesting to
compare with the case of \lmo, where below the transition the
$z^2$-orbital is $\sim65$\% filled, while above the transition the
angle $\theta$ amounted to $\theta\simeq\pi/2$ ($Q_3=0$), which also
corresponds to the equal filling of the $e_g$ orbitals
\cite{carvajal98}. In the later case the $Q_2$-type of the
octahedron distortions remain in the high-temperature phase due to
the interactions between the distortions. In our case the
interaction between the Jahn-Teller active $\rm Mn^{3+}$ ions is
weakened due to the dilution by 30\% of the statistically
distributed $\rm Mn^{4+}$ ions and thus the $\rm MnO_6$ octahedron
behaves like a free molecule with equally populated $e_g$-orbitals.
The difference in the in-plane bond lengths Mn-O2 in the OO state at
room temperature is an order of magnitude smaller in the case of
LPCM (cf. $0.27$~\AA\ vs. $0.02$~\AA). This abrupt decrease in the
bond length mismatch as a function of the $\rm Mn^{3+}$ fraction is
an additional evidence of the cooperative nature of the Jahn-Teller
effect. In the \lcm_ce\ where the ratio of $\rm Mn^{3+}:Mn^{4+}$ is
``50:50'' the $\rm MnO_6$-octahedron remains almost regular at room
temperature \cite{radaelli97}.

\subsection{Charge order transition}
\label{CO}

\begin{figure}[p]
\includegraphics[width=15cm]{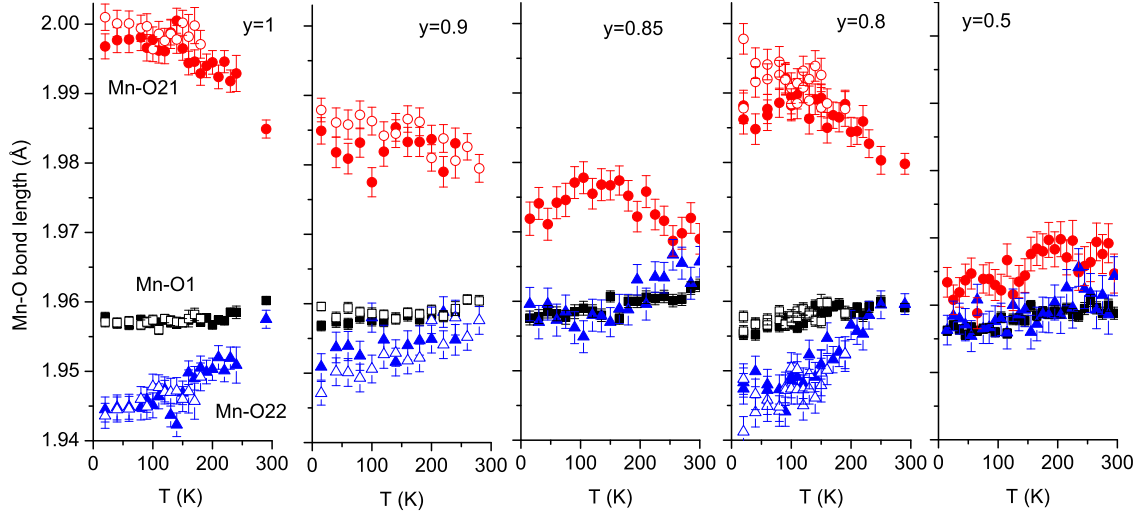}
\caption{(color online) Bond length as a function of temperature in \lpcm\ for
the Pr concentration y=1.0, 0.9, 0.85, 0.8, 0.5. The bond lengths for
the sample with y=0.7, are similar to ones for the y=0.8 and are shown
in Fig. \ref{70allT}.  \Os\ and \Oe-samples are shown by closed and
open symbols respectively.}
\label{disall}
\end{figure}

\begin{figure}[p]
\includegraphics[width=\figsize]{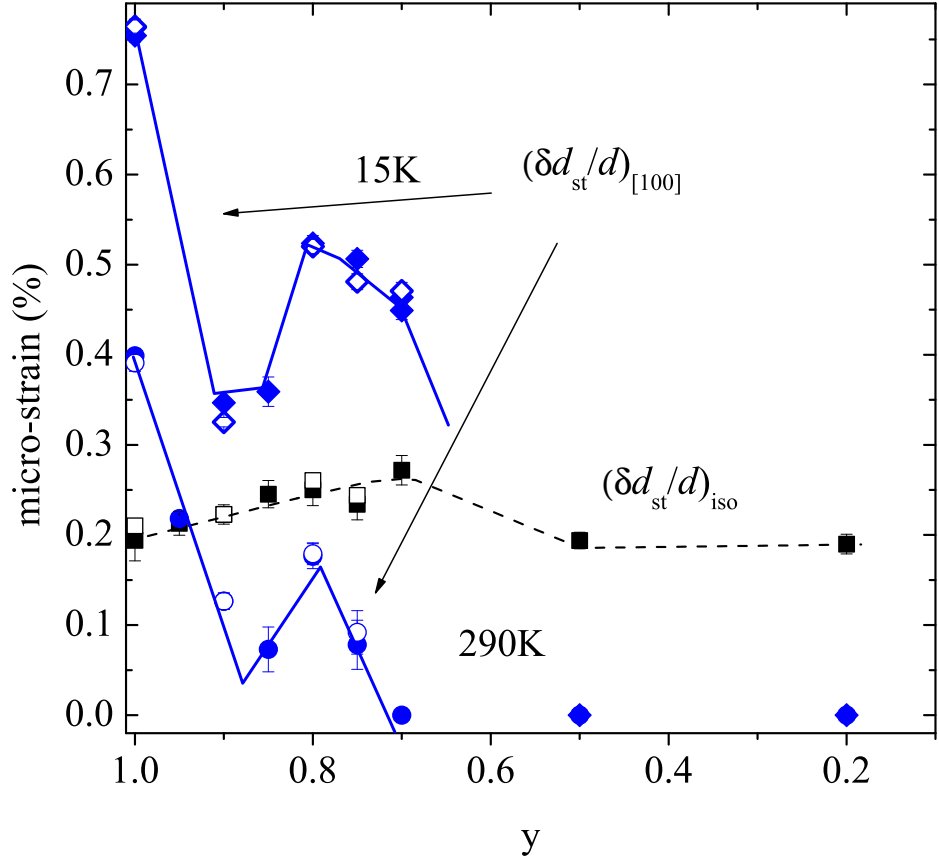}
\caption{(color online) (a) The lattice micro-strains as a function of Pr
concentration (sp. gr. $Pnma$) at $T=290$~K. The anisotropic
micro-strain $(\delta d_{st}/d)_{[100]}$ is also shown for
T=15~K. \Os\ and \Oe-samples are shown by closed and open symbols
respectively.}
\label{microstr}
\end{figure}

The LPCM system undergoes a transition to the charge ordered state
(CO) at $T_{CO}>T_N$ \cite{cox98,bala99:isotope}. The transition is
revealed by the appearance of weak superstructure peaks reflecting
the doubling of the unit cell along $a$-axis and the lowering of the
symmetry probably to the monoclinic $P2_1/m$ one similar to the case
of \lcm_ce~\cite{radaelli97}. However, the deviations from the
average $Pnma$ structure are small and do not allow a reliable
detection of the charge ordering. Better structure indicators of the
CO-transition are the increase in the Mn-O bond lengths mismatch and
the appearance of the anisotropic lattice micro-strains in the CO
state. One can see (Fig.~\ref{70allT}) that below $\sim250$~K the
two equal Mn-O bond length become different.  We believe that this
is a manifestation of the charge ordering which is schematically
shown as insert in Fig.~\ref{70allT}. This type of $\rm
Mn^{3+}/Mn^{4+}$ and orbital ordering in the $(ac)$ plane is
dictated by the Goodenogh-Kanamori rules and the low temperature
magnetic structure (PCE).  In Ref. \cite{cox98} there was proposed
that in \pcm\ the sites shown with the $z^2$-type orbital are fully
occupied by $\rm Mn^{3+}$, while those shown by circles are occupied
by $\rm Mn^{3+}:Mn^{4+}$ with the ratio $40:60$. Since the structure
is described in the average $Pnma$ space group the atom positions
are averaged over the symmetry elements of $Pnma$. The orbital
ordering in the charge ordered phase would give roughly equal
average in-plane Mn-O2 bond length. However, even in an ideal case
of CO in \lcm_ce\ the in-plane Mn-O2 bond lengths become different
in the CO-phase at low temperature in the
$Pnma$-description~\cite{radaelli97}.  The in-plane Mn-O2 bond
length mismatch in the LPCM is larger apparently due to the larger
$\rm Mn^{3+}$ fraction.  An evident indicator of the charge ordering
is the mismatch of the short Mn-O22 and Mn-O1 bond lengths. In the
antiferrodistorsive OO state all the $z^2$-orbitals are in the
$ac$-plane, while below the CO-transition the $z^2$-orbitals located
at ``$\rm Mn^{4+}$'' sites should be oriented perpendicular to the
$(ac)$-plane to stabilize the ferromagnetic coupling between the
neighboring planes along $b$-axis of the PCE-type antiferromagnetic
structure. This is reflected as a decrease in the short Mn-O22 with
respect to the Mn-O1 bond below $\sim250$~K (Fig.~\ref{70allT}).
Figure \ref{disall} shows the temperature dependences of the bond
lengths as a function of Pr-concentration $y$.  One can see that the
difference of the Mn-O1 and Mn-O22 bond lengths is drastically
decreased near $y=0.9$, reflecting the suppression of the charge
ordering. In the metallic predominantly ferromagnetic state
($y\leq0.5$) CO is suppressed as well. Another parameter which can
be considered as a structure indicator of the CO state is the
anisotropic micro-strain $(\delta d_{\rm st}/d)_{[100]}$ along
$a$-axis, along which the doubling of the unit cell occurs in the CO
phase. This implies that the CO/OO is not perfectly homogeneous as
in purely ``50:50'' composition \lcm_ce\, but there is a
distribution of domains with different extent of CO thus leading to
the dispersion of the lattice constant $a$. The anisotropic
micro-strain is pronouncedly increased on cooling for $y\geq0.7$,
but has a nonzero value at room temperature 290~K as shown in
Fig.~\ref{microstr}. A remarkable feature of $(\delta d_{\rm
st}/d)_{[100]}$ is a local minimum around $y_c=0.9$ pointing to a
suppression of the CO concomitantly with the suppression of the AFM
and FM ordering as discussed below. The isotropic micro-strain value
$(\delta d_{\rm st}/d)_{\rm iso}$ (also shown in
Fig.~\ref{microstr}) doesn't exhibit any peculiarity at $y_c$ ruling
out any kind of sample effect.

\subsection{Magnetic state}
\label{mag_state}

\begin{figure}[p]
\includegraphics[width=15cm]{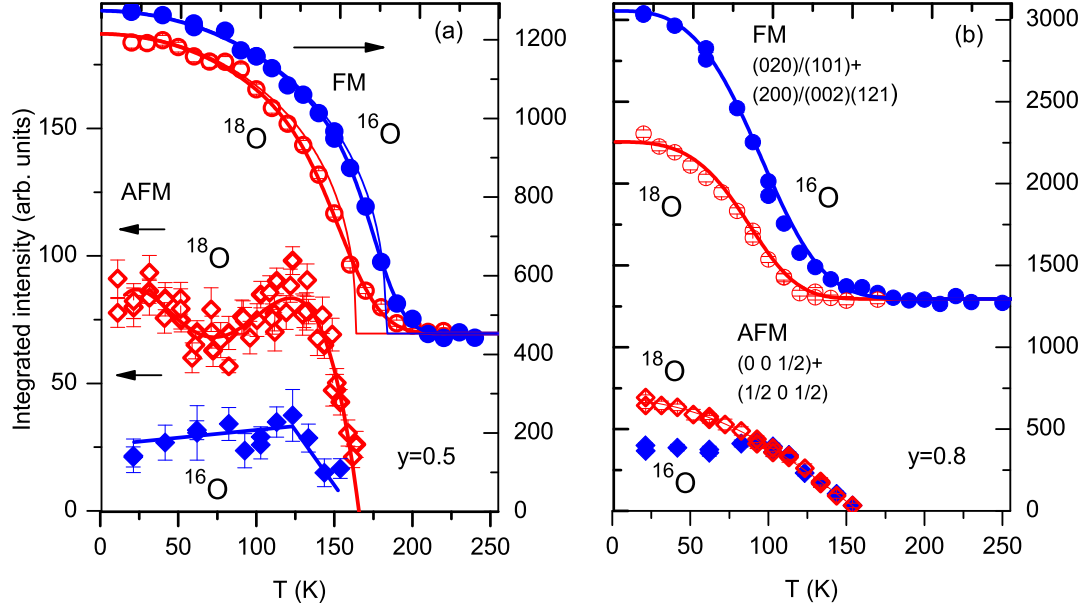}
\caption{(color online) Temperature dependences of the integrated intensities of the
selected magnetic diffraction peaks indicated in the
Fig. \ref{HRPTpatterns} in the samples with (a) $y=0.5$ and (b)
$y=0.8$. The data were collected on heating. \Os\ and \Oe-samples are
shown by closed and open symbols respectively.  The lines for the AFM
intensities are guides to the eye. For the FM ones the lines are fits to
the formula described in the text.}
\label{I50}
\end{figure}

The temperature dependences of the magnetic Bragg peak intensities
$I(T)$ allow accurate determining of the transition temperatures to
the AFM and FM states ($T_N$, $T_C$). A typical diffraction pattern
is shown in Fig. \ref{HRPTpatterns}, where the tick rows indicate
the nuclear and magnetic phases.  The magnetic structure consists of
AFM pseudo-CE component with Mn-spin directed along $b$ and FM
component in $(ac)$-plane similar to the one reported in
\cite{cox98, Bala01LPCM_magnetic}. In addition, the Pr-spins also
get ferromagnetically ordered parallel to the FM-component of
Mn-spins. The refined low temperature magnetic moments are presented
in Table~\ref{tab1}. Figure \ref{I50} shows several examples of the
temperature dependences of the selected magnetic Bragg peaks $I(T)$.
The \Os\ and \Oe\ samples for each $y$-concentration were prepared
in the same amounts and measured in the identical experimental
conditions and thus the integrated intensities shown in
Fig.~\ref{I50} were normalized only to the neutron monitor (number
of incoming neutrons). Above 100K the samples with $y\geq0.7$ are in
purely AFM state that is not affected by the oxygen substitution.
The Neel temperatures $T_N$ are the same and the whole sample volume
is occupied by the AFM phase, as illustrated by Fig.~\ref{I50}b. The
ferromagnetic phase is suppressed by the increase in the oxygen mass
that is reflected by the splitting of the AFM integrated intensities
for the \Oe\ and \Os\ samples roughly below 100K.

\begin{figure}[p]
\includegraphics[width=15cm]{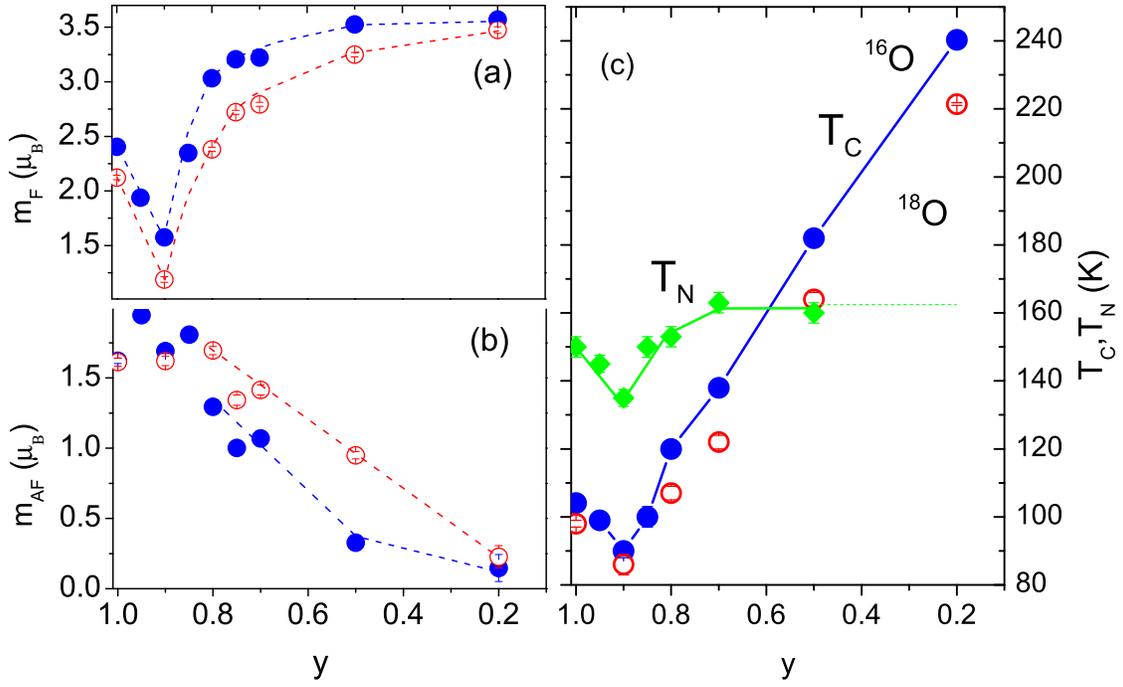}
\caption{(color online) Effective (a) ferro- $m_F$ and (b)
antiferromagnetic moments $m_{AF}$ at T=15~K, (c) N\'{e}el $T_N$ and
Curie $T_C$ transition temperatures determined from the ND data as a
function of $y$ in \lpcm. The lines are guides for the eyes. \Os\
and \Oe-samples are shown by closed and open symbols respectively.
N\'{e}el temperatures for the samples with different oxygen mass are
the same.} \label{T(y)}
\end{figure}

One sees that the AFM transition is very well defined by the abrupt
vanishing of the respective Bragg peaks. The ferromagnetic state is
destroyed in a more smoother way. Some fraction of the sample volume
remains ferromagnetic above the transition as one can see from the
smeared shape of the FM Bragg peak intensities $I(T)$ near $T_C$. To
determine the transition temperatures in an unbiased systematic way
we performed least-square fits of the experimental $I(T)$  in
assumption of the Gaussian distribution of the transition
temperatures and a phenomenological power-low function for the
spontaneous magnetization:

\beq{IofT}
I(T)=I_1 \int_{0}^\infty \frac{1}{\sqrt{2\pi}\delta
T_C}e^{(\tau-T_C)^2/2\delta T_C^2} (1-(T/\tau)^\alpha)^\beta
(1-H(T-\tau)) \,d\tau +I_0,
\eeq

where $H$ is a step-like Heaviside function, $T_C$ and $\delta T_C$
are the mean transition temperature and r.m.s. variance,
respectively. $I_1$ are $I_0$ are the refinable parameters. For the
compositions with $y>0.5$ the parameters $\alpha$ and $\beta$ were
fixed as indicated in Table~\ref{tab1}. Otherwise, the fit
parameters got unreasonable values. To reduce a number of the
refined parameters the experimental $I(T)$ curves for \Oe\ and
\Os-samples were processed in one combined fit with the same
parameters $\alpha$, $\beta$ and $I_0$. The results of the fits of
the ferromagnetic $I(I)$ to the formula (\ref{IofT}) are shown by
thick lines in Fig.~\ref{I50}. For comparison, the thin lines for
the sample with $y=0.5$ show the calculated $I(T)$ curves
deconvoluted from the Gaussian smearing. The magnetic transition
temperatures are summarized in the Table~\ref{tab1} and in Fig.
\ref{T(y)}. The Curie temperatures $T_C$ are in reasonable agreement
with the ones determined from the magnetic susceptibility data that
are also given in the Table \ref{tab1}. One can see from the
Fig.~\ref{T(y)} that the isotope substitution has a large effect on
the $T_C$, whereas the N\'{e}el temperature $T_N$ is invariant as
one would expect for the double exchange type of FM and the
superexchange AFM interactions, respectively.

The AFM phase has been detected at all concentrations, including
$y=0.2$. $T_N$ for the sample with $y=0.2$ was not determined,
because the AF-fraction is very small [the effective AFM moment at
$T=15$~K is $m_{A}=0.24(6)\mu_B$] and the temperature scan would
require extremely long measurements. For $y\geq0.7$, the AF ordering
precedes appearance of the FM phase with the temperature lowering.
The FM phase appears spatially separated, reducing the sample volume
occupied by the AFM phase and thus reducing the slope of the
$I_{AF}(T)$ roughly below $T_C$. The sizes of the FM- and
AFM-domains are mesoscopically large ($>10^3$~\AA) estimated in the
similar way as described in Ref.~\cite{Bala01LPCM_magnetic}. The
same effect of phase separation is present for $y=0.5$, where
$T_C>T_N$ (Fig \ref{I50}a) implying that there are two
characteristic energy scales connected with the ferromagnetism. One
($T_C$) is given by the double exchange mechanism with the strong
dependence on $y$. The second characteristic temperature is about
130~K where the phase redistribution in favor of the FM state is
triggered on.

\begin{figure}[p]
\includegraphics[width=\figsize]{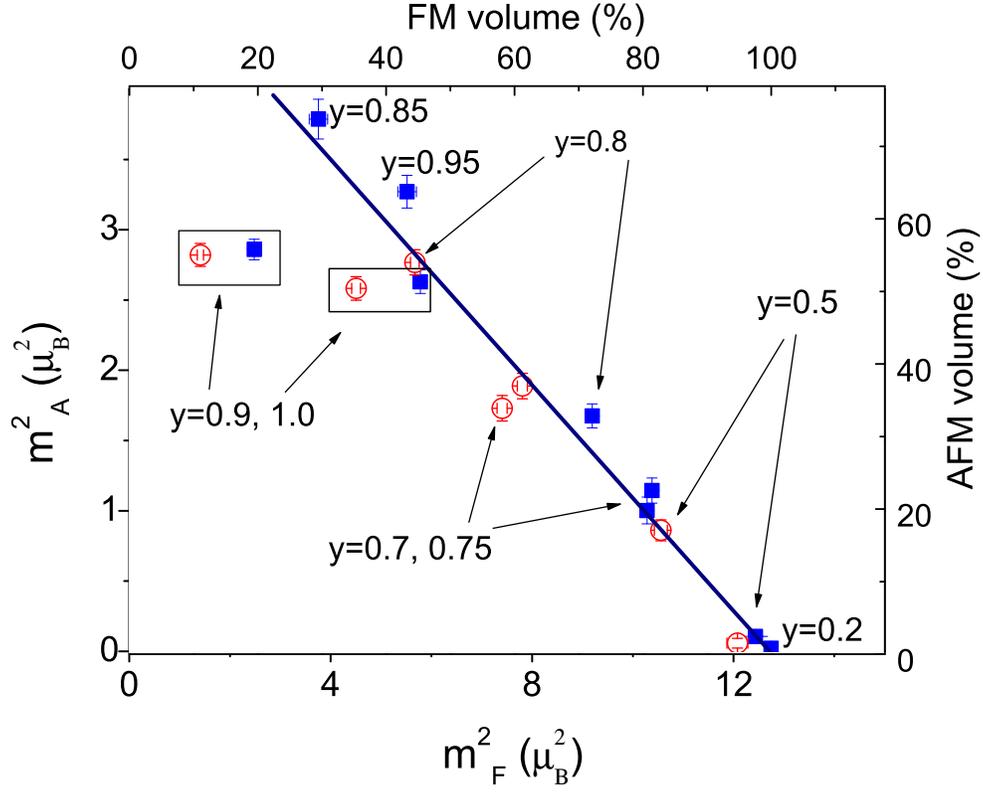}
\caption{(color online) Effective antiferromagnetic $m_{A}^2$ as a function of the
effective ferromagnetic moment $m_{F}^2$ for all the samples with $y$ as a
parameter indicated for each point. The top-$x$ and right-$y$ axes
show respective phase fractions. The line is a linear fit, y=0.9 and
y=1.0 (\Oe) are excluded from the fit. The closed and open symbols
represent the \Os\ and \Oe\ samples, respectively.}
\label{A(F)}
\end{figure}

Figure \ref{T(y)} shows the effective FM $m_F(y)$ and AFM $m_{A}(y)$
magnetic moments of Mn-ions at base temperature (10-15~K) as a
function of $y$. $m_A$ is the average value of the moments for two
sublattices $m_{A1}$ and $m_{A2}$ presented in Table~\ref{tab1}. The
effective moment is proportional to the product of the real moment
and square root of the volume occupied by the ordered phase. Since
the crystal structure is well refined in a single phase
approximation one cannot determine from a single diffraction
measurement whether the magnetic state is a homogeneous one.
However, it becomes possible from the analysis of the values of the
effective magnetic moments for all the concentrations $y$ for both
\Os\ and \Oe-samples.  Figure \ref{A(F)} shows $m_A^2(m_F^2)$ for
all the samples with the Pr-concentration $y$ as a parameter. The
line is a least-square fit to the linear function $m_A^2 = M_A^2
(1-{m_F^2/ M_F^2)}$ for all $y$ except $y=0.9$ and $y=1$ with \Oe.
This linear relation is virtually possible only in assumption of the
phase separated state. Otherwise, one would have to suppose quite
unreasonable sizes of the magnetic moment as a function of $y$ and
the oxygen mass. This linear function assumes that the magnetic
moments in the AFM and FM phases keep the constant values
$M_{A}=2.26(1)\mu_{B}$, $M_{F}=3.57(2)\mu_{B}$ over the whole phase
diagram, but the phase volumes are synchronously redistributed as
indicated by top-$x$ and right-$y$ axes of Fig.~\ref{A(F)}. The size
of the FM moment well corresponds to the average spin value of ${\rm
Mn^{3+}/Mn^{4+}}$ ions 3.7~$\mu_B$, whereas the AFM moment is
substantially lower. It can be connected with a ``defect'' charge
ordering that causes a frustration in the magnetic exchange between
the neighboring $(ac)$ planes.  With the help of the Fig. \ref{A(F)}
one can disentangle the effect of the FM/AFM volume redistribution
(the points slide along the straight line) and the effect of the
suppression of the magnetic ordering (the points lay below the
line). The increase in the oxygen mass simply shifts the balance
between the phases towards the AFM one for the ``metallic'' part of
the phase diagram ($y\leq0.8$), similar as the increase in $y$ does,
implying that the effect of the isotope substitution is the
polaronic narrowing of the bandwidth. The FM phase is metallic and
if its volume decreases below the percolation limit $\sim16$\%
\cite{gorkov04} then the sample should become an insulator.  In the
``insulating'' part of the diagram ($y\geq0.9$) the system behavior
is very different. Firstly, the ferromagnetic phase is not any more
metallic because its fraction, e.g. in the sample with $y=1$ is far
above a percolation threshold corroborating the insulating character
of the FMI phase. Secondly, the increase in the oxygen mass
suppresses only the FMI phase, both $T_C$ and the effective magnetic
moment, leaving the AFM phase fraction intact. One can speculate
that the FMI cannot be converted to the AFM phase because it is
already insulating and thus the further decrease in the electron
transfer integral doesn't necessarily shift the system towards the
AFM state. We suppose that those spins that stop contributing to the
FM phase become disordered thus reducing the effective moment.
However we cannot distinguish whether a spatially separated
disordered phase is formed or some spins inside the ferromagnetic
host get frustrated. The presence of the isotope effect on the FMI
is remarkable itself, because it rules out an ordinary superexchange
mechanism of FMI. This has been noted in Ref.~\cite{fisher03} where
the isotope effect on $T_C$ was found in insulating \pcmo, and the
genuine FMI state was proposed to appear due to some CO/OO
superstructure or density wave giving an energy gap which is much
smaller than the Hubbard on-cite repulsion.

\subsection{Effect of micro-strains on phase separation}
\label{u_str}

The specific low temperature FM/AFM volume fractions shown in Fig.
\ref{A(F)} do not represent themselves universal for LPCM system
values that depend only on $y$. Earlier we studied virtually the
same series of samples only with \Os
(N-series)\cite{Bala01LPCM_magnetic}, but prepared by the different
chemical route than the one explained in section \ref{exp}. The
phase redistribution process in the N-series samples (Fig.~8 of Ref.
\cite{Bala01LPCM_magnetic}) triggered at about $T_C$ on cooling goes
much ``easier'' leaving predominantly ($>90$\%) FM ground state for
$y\ge0.8$, while in the O-series samples the FM fraction is
decreased to 73\% for $y=0.8$.  The increase in the oxygen mass
leads to a complete phase volume redistribution in favor of the AFM
phase in the N-series sample with $y=0.75$~\cite{bala99:isotope}. In
the case of the O-series samples, the phase balance is also
definitely shifted towards the AFM state (Fig.~\ref{A(F)}), but none
of the \Oe-sample shows purely AFM ground state. We have repeated
the oxygen treatment and oxygen isotope exchange with the N-series
sample in the same way as we did with the samples of the present
O-series, including the back exchange and got essentially the same
results as in \cite{babushkina_nature, bala99:isotope}. It is
believed that the qualitative difference between the two series
originates from smaller grain sizes of the starting reagents in case
of the N-series that could give different micro-structure of the
sample crystallites.  We have undertaken a quantitative
micro-structure comparison of two exemplary samples with $y=0.75$
from each series also using ultra high resolution synchrotron x-ray
diffraction (as described in sec. \ref{crys_str}).  Both samples are
well refined with the close values of structure parameters. The
noticeable difference between the samples from the different series
is their micro-structure. The insert of Fig.~\ref{HRPTpatterns}
shows the $q$-dependence of the reduced peak broadening $\delta d/d$
in the two samples from both series. One can see that the broadening
is larger in the O-series sample. The main contribution to $\delta
d/d$ comes from the micro-strain effect amounted to $\delta d_{\rm
st}/d=0.16$\% and $0.23$\% for the N- and O-series sample
respectively. Inside the O-series the isotropic micro-strain varies
only slightly as shown in the Fig.~\ref{microstr}.  Thus, the key
difference between the two series of the LPCM-samples conditioning
their different ground states originates from the different values
of $(\delta d_{\rm st}/d)$ motivating to propose that the phase
separation is favored by the micro-strains. The value of the
micro-strains can be considered as a quantitative measure of the
disorder in the crystal lattice. Many studies of the epitaxial films
of manganites have demonstrated that the relative fractions of the
FM/AFM phases can be strongly modified by both morphology and
strains of the films (e.g.~\cite{wu03}).  The strong micro-strain
dependence of the phase separated volume fractions in our case
suggests that the value of the micro-strains $(\delta d_{st}/d)$ in
the bulk material should also be considered as an additional
parameter governing the formation of the phase separated state.

\subsection{Suppression of ordering near $y_c=0.9$}
\label{qcp}

The Curie temperature $T_C$ is decreased as a function of
Pr-concentration $y$ due to the smooth decrease in the Mn-O-Mn bond
angle (Fig.~\ref{r+angle}b), but near the critical concentration
$y_c=0.9$ one can see a local minimum (Fig.~\ref{T(y)}c), which is
unexpected from the behavior of crystal structure parameters. The
N\'{e}el temperature has a pronounced minimum at $y_c$, as well. The
total volume fraction occupied by both FM and AFM ordered phases
decreases by 25\% at $y_c$ in comparison with the other samples
(Fig.~ \ref{A(F)}). The charge ordering that is concomitant to the
antiferromagnetic PCE-type ordering is also suppressed near the
critical concentration $y_c$ (section~\ref{CO}).

The suppression of the magnetic transition temperatures  and the
ordered phase volumes at $y_c=0.9$ strongly resembles the phase
diagram calculated in Ref.\cite{burgy01}. The authors of Ref.
\cite{burgy01,burgy04} have shown that introducing of the quenched
disorder into an Ising system with two competing ordered states
gives an effect of a suppression of the ordering temperatures and
mesoscopic phase separation in the vicinity of the bi-critical
point. The experimental situation in our case is more complicated.
Instead of two phases there are three distinct phases: AFMI phase
and concomitant CO phase, FMM metallic phase for $y<y_c$ and FMI
insulating phase for $y>y_c$.  The phases, which would correspond to
the phases separated by the critical point are the FMM metallic from
one side and both FMI and AFM insulating phases from another side of
MI transition. In the spirit of the above model one would expect
that there is also a phase separation into the FMM and FMI phases
near $y_c$, which are indistinguishable from the present
experimental data.  The quenched disorder in the LPCM-system is
naturally present due the dispersion of the A-cation radius
amounting to $\sigma_A\simeq0.01$~\AA\ near $y_c$ (this value would
nominally correspond to quite large Pr-concentration spread $\Delta
y\simeq0.4$). However, since the radii of the Pr and Ca ions are
almost the same the dispersion $\sigma_A$ limits to zero for $y=1$,
while the phase separated state is still present. The role of the
quenched disorder can play the Jahn-Teller strain from $\rm Mn^{3+}$
centres as proposed in Ref.~\cite{argyriou04}. In addition, the
quenched disorder can originate from the chemically inhomogeneous
distribution of the A-cations given by the sample preparation
procedure.

We cannot exclude also that the lattice distortions and the
long-range strain \cite{littlewood99, ann04} can be the reason for
the phase separated state.  We note that the suppression effect was
not found in a detailed study of similar system \pcso\
\cite{blake02}, where all transition temperatures were continuous
through the critical point. The later system has almost the same
average A-cation radius but much larger ($\sim 4$ times) variance
$\sigma_A$, which could be a possible reason of the different
behavior. We suppose that to compare the different manganite systems
with respect to the effect of the suppression of the ordering near
$y_c$ one should also compare the extent of quenched disorder, which
can be quantitatively characterized by the value of the isotropic
micro-strains.

\section{Conclusions}

The magnetic and crystal structures have been studied in the series
of samples \lpcm\ as a function of the Pr-concentration ($y=$ 0.2,
0.5, 0.7, 0.75, 0.8, 0.85, 0.9, 0.95, 1.0) across the
metal-insulator (MI) transition around $y_c=0.9$ and as a function
of the oxygen mass (\Os, \Oe). The ground magnetic state is
mesoscopically separated into antiferromagnetic insulating (AFM) and
ferromagnetic (FM) phases for all values of $y$. The ordered
magnetic moments of the Mn-ions keep the constant values
$M_{AFM}=2.26(1)\mu_{B}$, $M_{FM}=3.57(2)\mu_{B}$ over the whole
phase diagram, whereas the phase fractions are changing as a
function of $y$.  The conductivity in the ``metallic'' part of the
phase diagram ($y<y_c$) is metallic because the FM phase is metallic
and its fraction is above the percolation threshold. In the
``insulating'' domain ($y>y_c$) the fraction of the FM phase is
significantly above the percolation threshold (e.g. for $y=1.0$ it
amounts to 46\%) implying that the FM phase is insulating for
$y>y_c$.

The increase in the oxygen mass has effect on both the values of
$T_C$ and the phase fractions.  The Curie temperatures of both
metallic and insulating FM phases are decreasing in a similar way
suggesting that the origin of the ferromagnetism in the insulating
state should be connected with a slow electron hopping process
similar to the double exchange as discussed in \cite{fisher03}. The
N\'{e}el temperature is not affected by the oxygen mass as expected
for the AFM superexchange. The increase in the oxygen mass always
suppress the FM phase fraction. However, the effect is different for
the metallic and insulating FM phase. For $y<y_c$, the oxygen
substitution acts similar to the increase in $y$, namely the FM
phase volume fraction is decreased by the same amount as the AFM
volume fraction is increased. For $y\ge y_c$ the ferromagnetic
insulating phase is suppressed but the AFM phase is not changed at
all. Thus, the narrowing of the bandwidth due to the decrease in the
A-cation radius and the polaronic narrowing act very differently in
the ``insulating'' part of the phase diagram.

The ratio between the AFM and FM phase fractions is not uniquely a
function of the effective bandwidth given by $y$ and the oxygen
mass, but is also affected by the value of internal micro-strains
that favor the formation of the phase separated state. We propose
that the value of the micro-strains can be used as a universal
quantity characterizing the extent of the quenched disorder, which
can have different origins in the manganites.

The phase diagram of \lpcm\ has a critical concentration at
$y_c=0.9$, where both N\'{e}el $T_N$ and Curie $T_C$ transition
temperatures, the ordered phase volumes as well as the charge
ordering concomitant to the AFM phase are suppressed. This finding
together with the presence of the phase separated state strongly
support the theoretical study of the competition between two ordered
states in the presence of quenched disorder\cite{burgy01,burgy04},
implying that the quenched disorder plays a key role in the
formation of the phase inhomogeneous state in manganites.

\section*{A{\lowercase{cknowledgements}}}

This study was partially performed at Swiss neutron spallation SINQ
and Swiss light source SLS of Paul Scherrer Institute PSI (Villigen,
PSI). We acknowledge A.~Kaul, O.~Gorbenko and N.~Babushkina for
granting us the sample LPCM75 we used in our previous study
\cite{bala99:isotope,Bala01LPCM_magnetic,Bala01LPCM_crystal},
L.~Keller for the help in neutron diffraction measurements. We thank
V.~Kabanov for the helpful discussions. The study was supported by the
RFBR (grant N06-02-16032).

\bibliography{publist,refs_manganites,refs_general}

\begin{table}[p]

\caption{The effective ferro- and antiferromagnetic moments at
$T\leq15$~K and the N\'{e}el $T_N$ and Curie $T_C$ transition
temperatures in the \LPCM. $T_C$ were determined from the fits of
neutron diffraction intensities to formula (\ref{IofT}) (ND) and by
the inflection point of magnetic susceptibility ($\chi$). $\delta
T_C$ is r.m.s. variance of the Gaussian distribution of the
transition temperatures. See the text for details. The values of the
moments are given in $\mu_B$. The Mn-magnetic moments of the
pseudo-CE AFM structure are $m_{A1}$ and $m_{A2}$ for the
propagation vectors $k_1$=[0 0 1/2] and $k_2$=[1/2 0 1/2],
respectively. $m_F$ is the FM moment of Mn-ion. $m_{\rm Pr}$ shows
the magnetic moment of Pr-ion. }

\footnotetext[1]{fixed} \footnotetext[2]{a=2 and b=0.66 are fixed in
formula (\ref{IofT}) } \label{tab1}
\begin{center}
\begin{tabular}{l|l|l|ll|l|l|l|l|l}

Oxygen      & $y$  & $T_N$ &$T_{C}$ ND;& $\chi$ &$\delta T_{C}$&$m_{A1}$ & $m_{A2}$ & $m_{F}$
                                    & $m_{\rm Pr}$                                        \\ \hline
\Os
            & 0.2  & -     &240.3(1);& 236(2) &0\footnotemark[1]  &0.06(18)    &0.23(5)       &3.57(2)
                                    & 0\footnotemark[1]     \\\hline
\Oe
            & 0.2  & -     &221.4(4);&224(2)  &0\footnotemark[1]  &0.23(8)     &0.26(7)       &3.46(3)
                                    & 0\footnotemark[1]     \\\hline\hline
\Os
            & 0.5  &160(3) &183(2);&180(2)    &14(2)              &0.36(5)     &0.29(7)       &3.53(2)
                                    & 0\footnotemark[1]     \\\hline
\Oe
            & 0.5  &160(3) &164(2);&155(2)    &19(2)              &0.95(3)     &0.91(4)       &3.25(2)
                                    & 0\footnotemark[1]     \\\hline\hline
\Os   \footnotemark[2]
            & 0.7  &163(3) &138(2);&122(2)    &30(3)              &1.11(5)     &1.03(4)       &3.22(2)
                                    & 0.48(2) \\\hline
\Oe   \footnotemark[2]
            & 0.7  &163(3) &122(2);&116(2)    &29(3)              &1.41(4)     &1.33(3)       &2.79(2)
                                    & 0.64(3) \\\hline\hline

\Os
            & 0.75 &-      &-     ;&122(2)    &-                  &1.03(5)     &0.97(5)       &3.21(2)
                                    & 0.52(2) \\\hline
\Oe
            & 0.75 &-      &-     ;&116(2)    &-                  &1.34(4)     &1.29(3)       &2.72(2)
                                    & 0.63(3) \\\hline\hline

\Os   \footnotemark[2]
            & 0.8  &153(3) &120(1);&120(2)    &23(2)              &1.30(4)     &1.28(3)       &3.03(2)
                                    & 0.49(2) \\\hline
\Oe   \footnotemark[2]
            & 0.8  &153(3) &107(2);&114(2)    &16(3)              &1.63(3)     &1.66(2)       &2.38(2)
                                    & 0.57(3) \\\hline\hline
\Os   \footnotemark[2]
            & 0.85 &150(3) &100(3);&116(2)    &35(4)              &1.82(3)     &1.79(3)       &2.34(4)
                                    & 0.35(6) \\\hline\hline
\Os   \footnotemark[2]
            & 0.9  &135(3) & 90(2); &99(2)    &21(3)              &1.63(3)     &1.75(2)       &1.58(2)
                                    & 0.39(3) \\\hline
\Oe   \footnotemark[2]
            & 0.9  &135(3) & 86(3); &94(2)    &17(4)              &1.62(3)     &1.74(2)       &1.19(3)
                                    & 0.63(4) \\\hline\hline
\Os   \footnotemark[2]
            & 0.95 &145(3) &99.0(9);&117(2)    &17(1)              &1.96(3)     &1.93(4)       &1.94(5)
                                    & 0.39(7) \\\hline\hline
\Os    \footnotemark[2]
            & 1.0  &150(3) &104.1(8);&120(2)    &14(2)              &1.63(3)     &1.61(3)       &2.40(2)
                                    & 0.53(2) \\\hline
\Oe    \footnotemark[2]
            & 1.0  &150(3) & 98(1);&115(2)    &15(2)              &1.62(3)     &1.60(3)       &2.12(2)
                                    & 0.61(2) \\\hline\hline

\end{tabular}

\end{center}
\end{table}


\end{document}